\documentclass[sigconf]{acmart}
\usepackage{cleveref}
\usepackage{subfig}
\usepackage{textcmds}
\AtBeginDocument{%
  \providecommand\BibTeX{{%
    \normalfont B\kern-0.5em{\scshape i\kern-0.25em b}\kern-0.8em\TeX}}}

\copyrightyear{2022} 
\acmYear{2022} 
\setcopyright{acmlicensed}\acmConference[MSR '22]{19th International Conference on Mining Software Repositories}{May 23--24, 2022}{Pittsburgh, PA, USA}
\acmBooktitle{19th International Conference on Mining Software Repositories (MSR '22), May 23--24, 2022, Pittsburgh, PA, USA}
\acmPrice{15.00}
\acmDOI{10.1145/3524842.3528524}
\acmISBN{978-1-4503-9303-4/22/05}

\begin{document}

\title{Replicating Data Pipelines with GrimoireLab}

\author{Kalvin Eng, Hareem Sahar}
\email{{kalvin.eng, hareeme}@ualberta.ca}
\affiliation{%
  \institution{University of Alberta}
  \city{Edmonton}
  \country{Canada}
}








\renewcommand{\shortauthors}{Eng and Sahar}

\begin{abstract}
In this paper, we present our MSR Hackathon 2022 project that replicates an existing Gitter study~\cite{sahar2021issue} using GrimoireLab. We compare the previous study's pipeline with our GrimoireLab implementation in terms of speed, data consistency, organization, and the learning curve to get started. We believe our experience with GrimoireLab can help future researchers in making the right choice while implementing their data pipelines over Gitter and Github data.

\end{abstract}



\maketitle

\section{Introduction}
Developer chat rooms such as Slack and Gitter are frequently used for project specific discussions. These discussions contain a wealth of information that can be leveraged to facilitate better software development and management. Gitter is a chat platform that is mainly centered around Github repositories which can offer a wealth of information such as informal discussions centered around issues of a project. The mining of this data surrounding a project is a non-trivial task as it spans through Github and Gitter with disparate APIs and schemas.

As such, tools like GrimoireLab~\cite{duenas2021grimoirelab} have been developed to support tasks such as data retrieval, processing, and visualization from software repositories, and other relevant sources. Consequently, researchers and analysts can retrieve large datasets in an efficient way, without reinventing the wheel, and at the same time ensuring easier replicability of their work.

The primary goal of our hackathon project was to replicate the data pipeline of an existing study by \citet{sahar2021issue} that investigates chat discussions among developers in the Gitter platform. \citet{sahar2021issue} investigated: how frequently issues are mentioned in Gitter chat; reasons why issues are posted in chat; whether or not posting issues in chat affects their resolution time; and whether mentioning issues in chat are correlated with  number of Github issue comments. We do not fully replicate their study as some analysis steps involved manual labelling, rather we investigate whether the use of GrimoireLab makes the data collection process easier. To this end, we use several GrimoireLab components to replicate the pipeline and answer some of the data distribution related questions from the original paper. 

\section{Previous Data Pipeline}
\begin{figure}[h]
  \centering
  \includegraphics[width=0.75\linewidth]{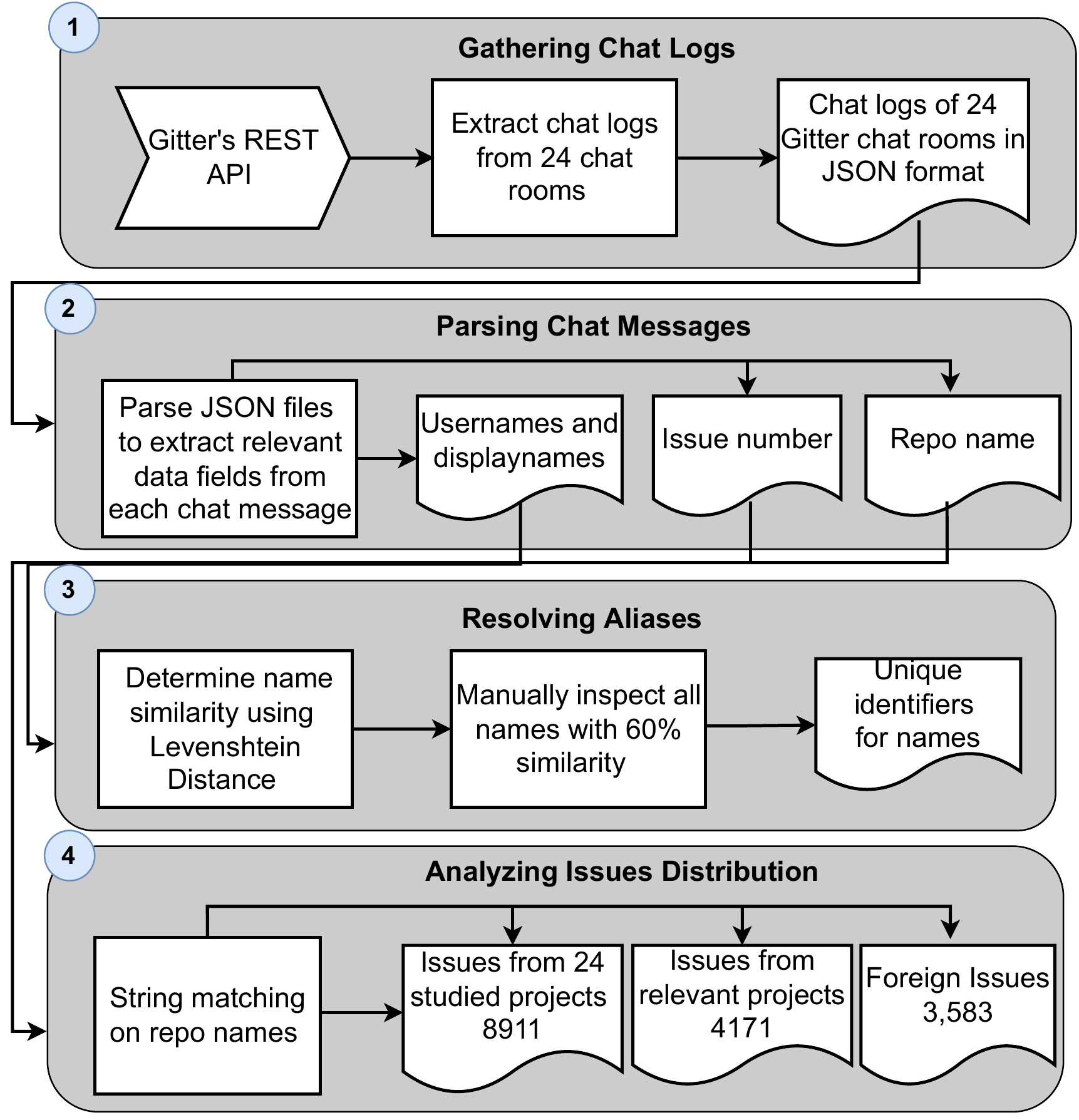}
  \caption{Methodology of \citet{sahar2021issue} pipeline.}
  \label{sahar-pipeline}
  \Description{Methodology of previous pipeline.}
\end{figure}

\citet{sahar2021issue} gathered data from 24 Github repositories 
and their associated Gitter chat rooms. 
\Cref{sahar-pipeline} shows their data pipeline:
\begin{enumerate}
    \item \textbf{Data Retrieval} - The Gitter API was used to extract chat logs. The chat logs, which were in JSON format, were parsed to obtain issue references. The issues and their metadata was extracted from Github using the Github API. 
    \item \textbf{Data Storage} - The chat logs were stored in JSON whereas issues were stored in CSV files.
    \item \textbf{Identities Management} - The Gitter and Github username and displayName are compared to resolve aliases. For this, Levenshtein distance was used and later manual analysis was done.
    \item \textbf{Analytics} - An analysis was done on the aligned and cleaned data to answer four research questions, one of which involved open coding to manually label purpose of issue references in Gitter.
\end{enumerate}

The data pipeline was implemented using a mix of R, Python, and bash scripts. The dataset and scripts of this process can be found on Github.~\footnote{\url{https://github.com/Hareem-E-Sahar/gitter}}

\section{GrimoireLab Pipeline}
\begin{figure}[h]
  \centering
  \includegraphics[width=\linewidth]{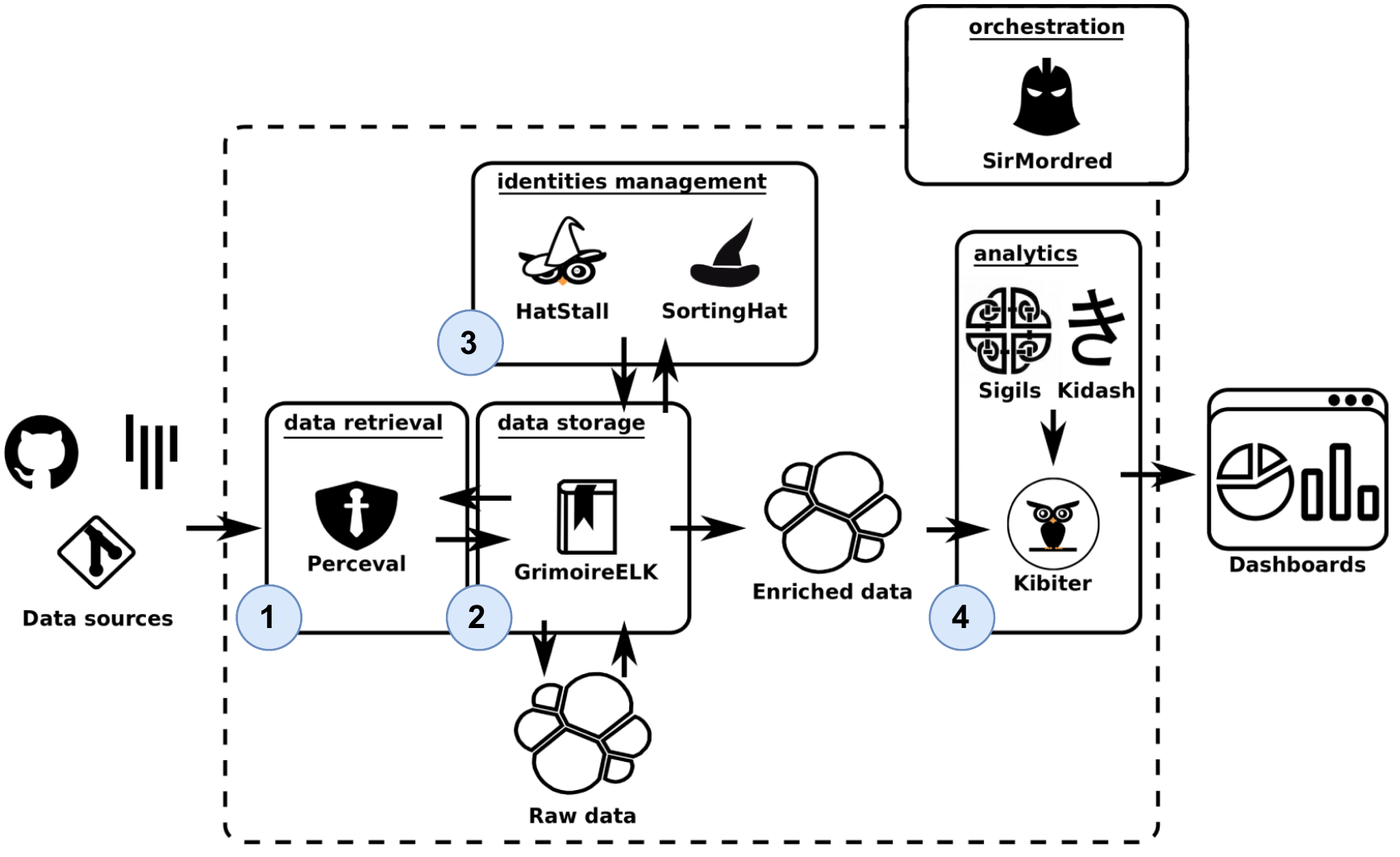}
  \caption{Components of GrimoireLab used in our study.}
  \label{griomirelab-pipeline}
  \Description{Components of GrimoireLab used.}
\end{figure}
\vspace{-3mm}

To use GrimoireLab, we choose the Docker Compose approach which orchestrates all the relevant Docker containers needed to work with GrimoireLab. This approach allows users to deploy GrimoireLab without installing any dependencies on the host machine. 

In particular, the SirMordred container contains all the tools needed for data retrieval, data storage, identity management, and managing dashboards. MariaDB and ElasticSearch containers are used for data storage in the form of inverted indexes, whereas the Kibiter container is used to create dashboards and to visualize the collected data. 
 \Cref{griomirelab-pipeline} shows the GrimoireLab components used in this work:

\begin{enumerate}
    \item \textbf{Data Retrieval} - Raw data is collected via \textit{Perceval} from the Github API, Gitter API, and Github git repositories and inserted into an ElasticSearch index.
    \item \textbf{Data Storage} - \textit{GrimoireElk} enriches the raw data scraped in the previous step, e.g., identifying the issues and pull requests in a Gitter message and assigning identities determined in (3).
    \item \textbf{Identities Management} - \textit{SortingHat} helps to manage identities retrieved from different data sources and allows for similar identities to be merged together. \textit{HatStall} is used as the user web interface for viewing and merging the identities.
    \item \textbf{Analytics} - \textit{Kibiter} is a fork of the Kibana dashboard that helps to creates data visualizations from ElasticSearch indices. To enhance \textit{Kibiter}, premade dashboard panels called \textit{Sigils} are used. For backing up and loading panels, \textit{Kidash} is used.
\end{enumerate}
\textit{SirMordred} serves as the main program to orchestrate task execution among the components.
The scripts used to generate the dataset can be found on Github.~\footnote{\url{https://github.com/k----n/GrimoireGitter}}

\subsection{Adaptations}
In order to replicate the results of the previous pipeline, some adaptations were needed to made for GrimoireLab to enrich the Gitter data correctly and align Github and Gitter identities.

\paragraph{\textbf{Gitter Data Enrichment}} We found that \textit{GrimoireElk}, which enriches the raw data of Gitter, did not accurately classify issues and pull requests in Gitter messages similar to the previous pipeline implementation. Therefore, we adapt \textit{GrimoireElk} to identify the issues and pull requests from the Gitter raw data via querying Github. In line with \citet{sahar2021issue}, we also classify the source repository of the issue and pull requests into 3 categories: \textit{project} (repository is directly related to the Gitter room), \textit{parent} (repository is related to the parent organization of the current project discussion in Gitter room), and \textit{foreign} (has no relation to the parent organization or repository in the Gitter Room).

\vspace{-2.5mm}
\paragraph{\textbf{Identity Alignment}} Although \textit{SortingHat} provides matching techniques for identities, we find that it can be insufficient for identities that might have similar names or usernames. Therefore, we implement a detection script that queries the \textit{SortingHat} identities that have not been merged yet and find identities that have a normalized Levenshtein distance greater than 0.7 for both names and usernames. The detected pairs of identities can then be merged using the \textit{HatStall} web interface if the end user decides that they are similar enough. 

\vspace{-2.5mm}
\section{Preliminary Results}
Using GrimoireLab we replicate the previous pipeline on 7 out of the 24 Gitter rooms\footnote{The data of 4 Gitter rooms used in the previous study are no longer available.}: 
\href{https://gitter.im/amberframework/amber}{\textit{amberframework/amber}}, 
\href{https://gitter.im/aws/aws-sdk-go}{\textit{aws/aws-sdk-go}}, 
\href{https://gitter.im/patchthecode/JTAppleCalendar}{\textit{patchthecode/JTAppleCalendar}},
\href{https://gitter.im/mailboxer/mailboxer}{\textit{mailboxer/mailboxer}}, \href{https://gitter.im/PerfectlySoft/Perfect}{\textit{PerfectlySoft/Pe\-rfect}}, 
\href{https://gitter.im/kriasoft/react-starter-kit}{\textit{kriasoft/react-starter-kit}}, and 
\href{https://gitter.im/shuup/shuup}{\textit{shuup/shuup}}. 

\begin{figure}[h]
  \centering
  \includegraphics[width=0.8\linewidth]{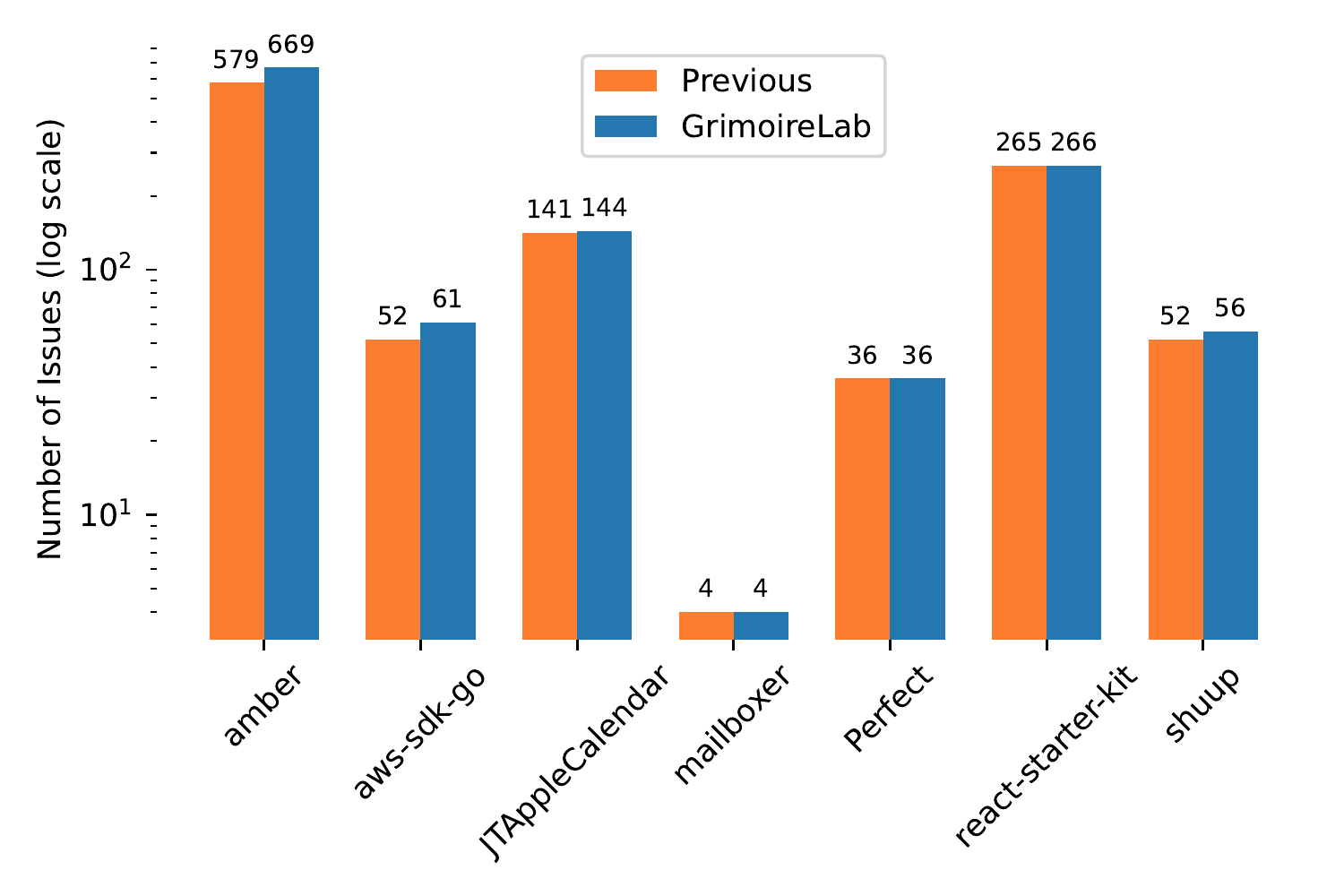}
  \vspace{-5.5mm}
  \caption{Count of Gitter API issues.}
  \label{api-counts}
\end{figure}

\Cref{api-counts} shows the issue references found by \citet{sahar2021issue} to the ones we found via GrimoireLab. We retrieved a more issues using GrimoireLab per project as it collects data until February 2022 whereas original study used data until November 2019. For our comparisons, we use data from both pipelines up until November 2019.

\vspace{-5.5mm}
\vspace{-2.5mm}

\begin{figure}[tbp]
  \centering
  \includegraphics[width=0.75\linewidth]{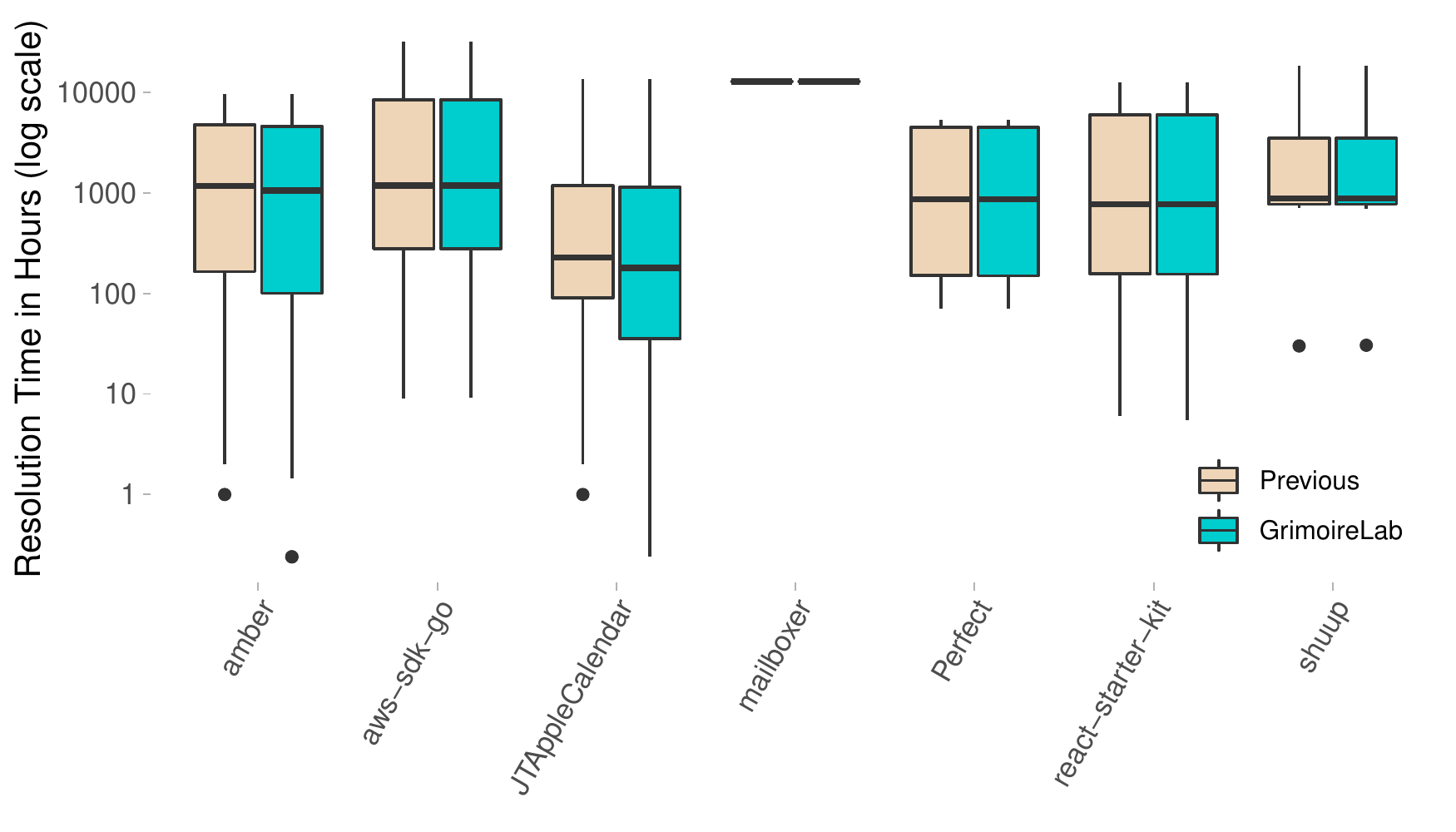}
  \vspace{-5mm}
  \caption{Resolution time comparison between previous pipeline and GrimoireLab.}
   \label{fig:resolution-times} 
\end{figure}

We compare the resolution time of issues and pull requests referenced in Gitter rooms as shown in \Cref{fig:resolution-times}.
We can see that the boxplots are relatively similar with some minor differences which is attributed to the different second-precision for the calculated resolution times.

\begin{figure}[tbp]
  \centering
  \includegraphics[width=0.75\linewidth]{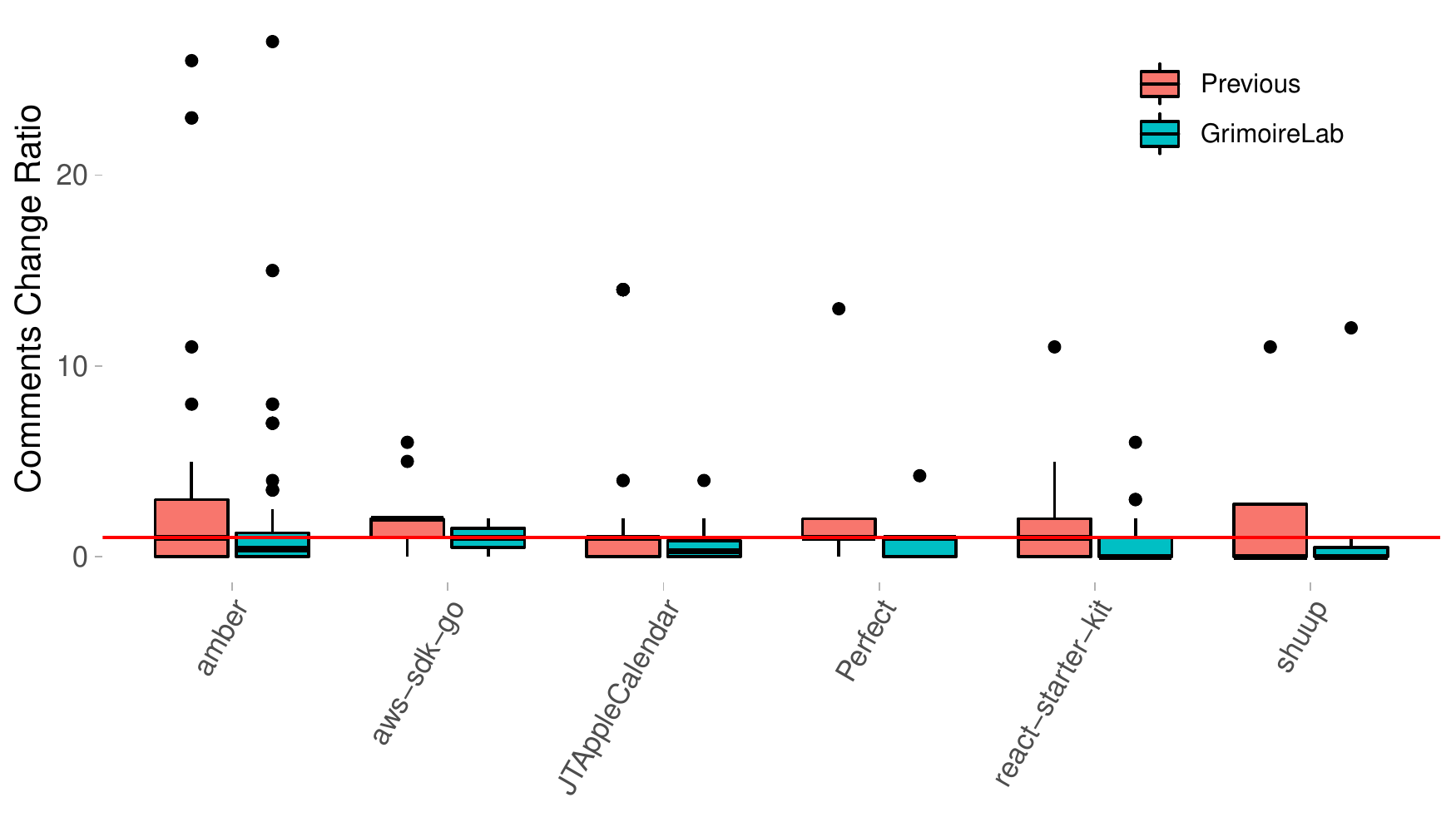}
  \vspace{-5mm}
  \caption{Ratio of number of issue comments in Github one week after and before issue reference in Gitter.}
  \vspace{-5mm}
  \label{comment}
\end{figure}

In \Cref{comment}, we show the comments change ratio which is computed by dividing the number of comments found in the Github issue tracker before being referenced in Gitter and by the number after within one week. The previous study did not include the comments posted on the day that the Gitter issue reference was made. As a result, their results were different than ours which can be seen from the median of boxplots and the outliers. GrimoireLab precisely computes the dates of comments to a second level granularity which allowed us to compute results with more precision than the previous study~\cite{sahar2021issue}.
    \hfill

\section{Comparison of Approaches}
We anecdotally compare the pros and cons of adopting a traditional pipeline approach \cite{sahar2021issue} to the GrimoireLab pipeline in terms of speed, data consistency, organization, and the learning curve.

\paragraph{\textbf{Speed}}
We only choose 7 Github repositories for the replication since they contain at most 4214 issues and pull requests (as of February 3, 2022) making it manageable to be scraped from Github. Large Github projects are difficult to scrape with GrimoireLab as it is designed to scrape all data from Github which requires many tokens\footnote{Github API allows 5,000 requests per hour against each token.}. With a few tokens, the time to retrieve large Github projects with issues and pull requests in the tens of thousands would likely take days to retrieve. In comparison, \citet{sahar2021issue} extracted only specific issues and pull requests mentioned in Gitter, which are relatively small in number. As a result they were less limited by the API rate limits and could obtain data from large projects relatively quickly.

In terms of data processing speed, the GrimoireLab data enrichment process can take a long time due to the sheer volume that it must process --- it took approximately 3 hours to create complete enriched indices for the 7 repositories. Comparatively, the previous pipeline can be more selective in data processing, hence it could be significantly faster.

\vspace{-1.5mm}
\paragraph{\textbf{Data Consistency and Organization}}
We find that GrimoireLab helps to keep data more organized out of the box by providing tools to centralize data into ElasticSearch, as well as manage identities in a MariaDB database. All of the collected and processed data can be accessed via a single ElasticSearch API. Moreover, identities can be easily managed via a web interface with \textit{HatStall} or via the command line with \textit{SortingHat}.

By comparison, the previous pipeline used a collection of CSV and JSON files to store the retrieved and processed data. Multiple scripts were used to extract data, align identities, remove inconsistencies, and finally compute results.
Having a non-standardized collection of files with differing schemas makes querying for data more difficult and could possibly lead to inconsistencies. This became very clear when we computed the after and before comments ratio in \Cref{comment} and found differing results. Furthermore, we also found that resolution times in \Cref{fig:resolution-times} were not as precise in the previous pipeline compared to the GrimoireLab data likely due to dates in the previous pipeline being processed through multiple scripts. We found that GrimoireLab produced more accurate data from a single source, as opposed to the data being \qq{lost in translation} in the previous pipeline from using multiple scripts and files to produce the data.

\vspace{-1.5mm}
\paragraph{\textbf{Learning Curve}}
GrimoireLab is a diverse collection of tools separated into microservices, so it took us a lot of time to become familiar with the software stack. We had to read extensive documentation and source code to understand the particular use cases of the GrimoireLab components. Furthermore, familiarity is also required in software such as Docker to bootstrap the pipeline quickly without the need to install software dependencies. ElasticSearch familiarity is also needed to help query the retrieved data in GrimoireLab. Learning the different schemas of ElasticSearch indices created with GrimoireLab can be quite daunting, but once familiarized with the indices, powerful visualizations can be created in the \textit{Kibiter} dashboard to explore data.

In comparison, the previous pipeline's implementation is a collection scripts that requires no need for orchestration. Instead, R, Python, and bash scripts and executed in a consequential manner which makes the implementation much simpler. However, once the number of projects to analyze grow much larger, we anticipate that GrimoireLab will be much easier to work with as there is already built-in support for retrieving data from multiple sources and a singular API for querying the data. We confirmed this by retrieving and processing issue comments, which we found was easier with GrimoireLab.

\section{Conclusion}
In this paper, we describe our experience with using GrimoireLab to implement an existing pipeline~\citet{sahar2021issue}. Our preliminary results indicate that the GrimoireLab pipeline is superior compared to a traditional pipeline in terms of data consistency and organization, but it offers a steep learning curve. 
A complete replication package of our GrimoireLab usage can be found on Github here: \url{https://github.com/k----n/GrimoireGitter}

\bibliographystyle{ACM-Reference-Format}
\bibliography{main}










\end{document}